\documentclass[aps,reprint,groupedaddress, amsmath]{revtex4-1}

\usepackage{graphicx}
\usepackage{bm}

\begin{document}

\title{Attospiral generation upon interaction of circularly polarized intense laser pulses \\
with cone-like targets }

\author{Zs. L\'ecz}
\email{zsolt.lecz@eli-alps.hu}
\affiliation{ELI-ALPS, ELI-HU NKft. Dugonics square 13., 6720 Szeged, Hungary}

\author{A. Andreev}

\affiliation{Max-Born Institute, Berlin, Germany}
\affiliation{ELI-ALPS, ELI-HU NKft. Dugonics square 13., 6720 Szeged, Hungary}

\date{\today}

\begin{abstract}

Generation of high intensity attopulses is investigated in cylindrical geometry by using 3D particle-in-cell plasma simulation code. Due to the rotation symmetric target, a circularly polarized laser pulse is considered propagating on the axis of a hollow cone-like target. The large incidence angle and constant ponderomotive pressure leads to nano-bunching of relativistic electrons responsible for the laser-driven synchrotron emission. A numerical method is developed to find the source and direction of the coherent radiation that is responsible for the existence of attopulses. The intensity modulation in the harmonic spectrum is well described by the model of coherent synchrotron emission extended to the regime of higher order $\gamma$-spikes. The spatial distribution of the higher harmonics resembles a spiral shape which gets focused into a small volume behind the target.
\end{abstract}

\maketitle

The theory of surface high harmonic generation (SHHG) on the boundary of an overdense plasma was developed and well established a decade ago \cite{bulanov_theory, shortpulse, theory,HHG_relat, CWE}. Two main mechanisms are responsible for the generation of higher harmonics in overdense plasma: the coherent wake emission and reflection from a relativistically oscillating mirror (ROM) \cite{plasmamirror}, which is more efficient at highly relativistic intensities. A weak transition between them is observable at non-relativistic laser intensities by varying the plasma scale length \cite{transition}. In the case of near-critical plasma isolated attopulses can be produced by the effect of phase-dependent plasma mirror deflection \cite{attopulses}. The universal spectrum proposed for the ROM spectrum in Ref. \cite{theory} found to be invalid in some cases of oblique incidence or for plasma boundaries possessing finite density scale length \cite{Boyd2008, Pukhov2009}. Few years later a new mechanism was identified providing more intense attopulses via electron nanobunching and coherent synchrotron emission (CSE) \cite{pop_pukhov}. Generally, the SHHG in the relativistic laser plasma interaction can be interpreted as laser-driven synchrotron emission \cite{Mikhailova} by electrons moving collectively in the laser field near the plasma boundary.

The CSE is analytically described in one dimension and it is found to be very sensitive on laser and plasma parameters, which makes difficult the experimental realization in a controlled way. However, electron nanobunches are observed in 2D simulations for large incidence angles ($>70$ degrees) and $p-$polarization \cite{andreev, attobunches}. In that case the linearly polarized intense laser light can be efficiently converted into attopulses with the same polarization. In the present work we study this attopulse generation process in 3D in temporal and spatial domains and we show that the measured frequency spectrum is well described by numerically evaluated CSE spectra in the regime of higher order gamma-spikes \cite{pukhov}. Instead of a flat foil and linearly polarized pulse we use cylindrical targets and circularly polarized (CP) pulse in order to ensure conditions for p-polarized incidence at each point of the inner target surface. This setup automatically results in focusing of the generated radiation, which is a different approach for achieving higher attopulse intensities \cite{atto_prop, CHF}. The generation of ultraviolet vortices \cite{HHG_XrayVortice} has been recently proven using higher order Laguerre-Gaussian modes of the laser pulse. Here we show that similar attopulse-vortex (or attospirals) can be produced with much higher energy conversion efficiency via interaction of CP pulse with overdense plasma. Such pulses, carrying orbital angular momentum \cite{HHG_XrayVortice}, can be used for diagnostics in turbulent plasmas \cite{Mendonca}, detection of electron vortices \cite{electron_vortex} or structural exploration of materials by polarization-dependent absorption spectroscopy \cite{photonics}.

Modulations in the spectral intensity of emitted radiation has been observed in experiments \cite{IWatts} and simulations \cite{Boyd2008, Pukhov2009}, but in both cases these modulations appear as irregular fluctuations, therefore it has never been compared to analytical spectra. Here we present, for the first time, a direct comparison between the spectrum of coherently emitted radiation and analytical CSE spectrum. The most attractive feature of this type of spectrum is the constant intensity up to a roll-over frequency, which increases with laser intensity, but it is lower than the value predicted by the ROM model \cite{Pukhov2009}. Above this frequency the spectrum has an exponential decay modulated by periodic structures.

The simulations presented in this paper were performed in VORPAL (VSim) 3D. The interaction of a few cycle CP pulse with cylinder and cone targets is considered. The laser pulse propagates along the axis of the targets, thus the case of cylinder corresponds to tangential, while the cone target to grazing incidence. For the cone target $\alpha=$12 degrees half opening angle with respect to the longitudinal axis is considered. The cylinder radius is $R_t=0.8\,\mu$m with wall thickness of $d=0.1\,\mu$m, which is in the non-transparent regime because the plasma skin depth is
$l_p = a_0\lambda_L n_{cr}/(\pi n_0)<$~50~nm \cite{thinfoil}. The total length of the target is 2 $\mu$m and its density is $n_0=28n_{cr}$. The laser pulse is 20 fs long (10 fs FWHM) with Gaussian temporal profile and with a peak intensity $I_L=10^{20} W/cm^2$ corresponding to a normalized peak amplitude of $a_0=(I_L[10^{18} W/cm^2]\lambda_L^2[\mu m^2]/1.37)^{1/2}=6.8$. The laser wavelength is $\lambda_L=0.8\,\mu$m and the focal spot size of the Gaussian pulse (FWHM) is 2 $\mu$m. The grid size used in the simulation is $dx=$10 nm in each directions (see supplemental material \cite{supplem} for detailed simulation configuration).

\begin{figure}[h]
\centering
\includegraphics[width=42mm]{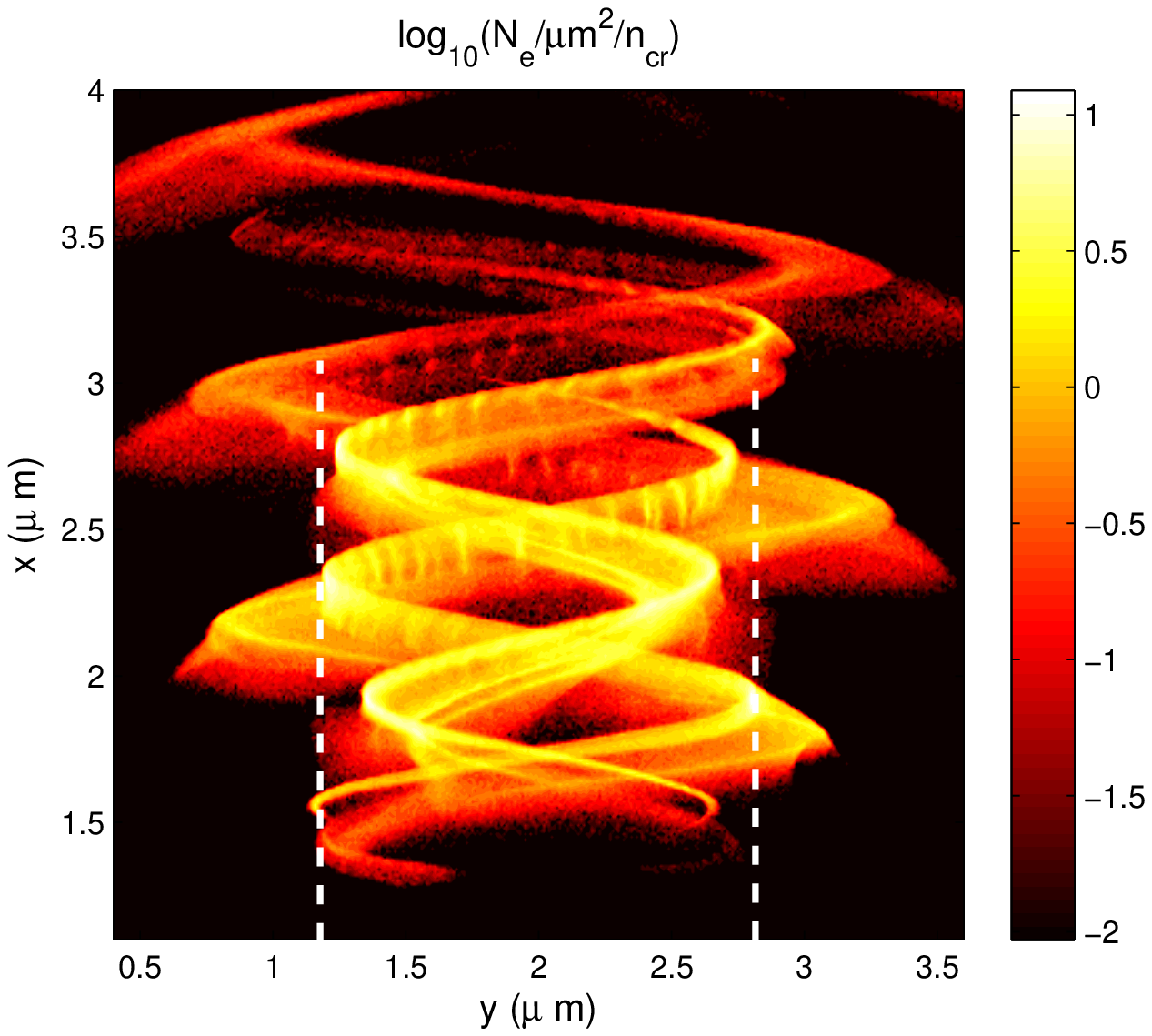} \hspace{-5mm}
\includegraphics[width=44mm]{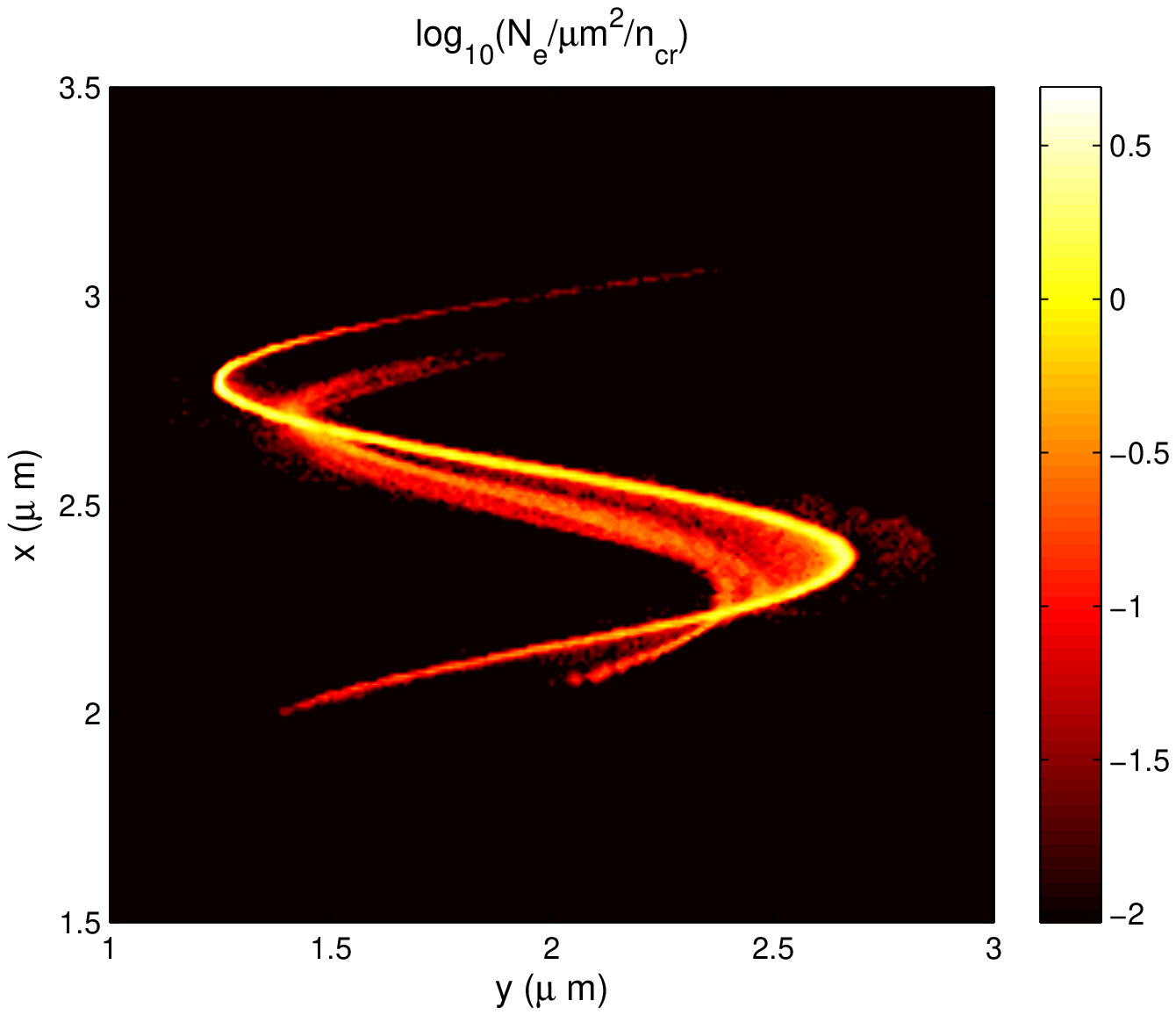}
\includegraphics[width=42mm]{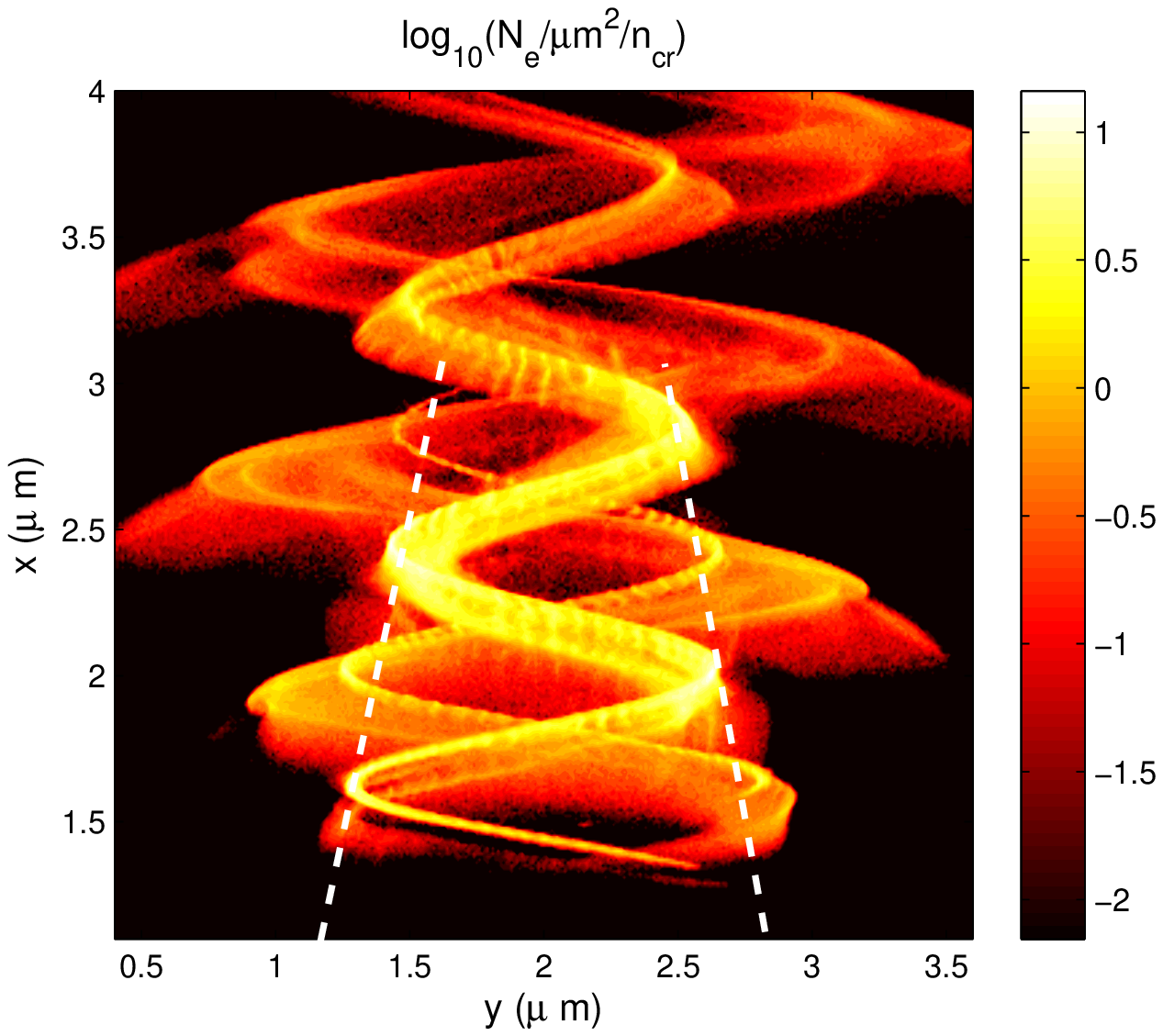} \hspace{-5mm}
\includegraphics[width=44mm]{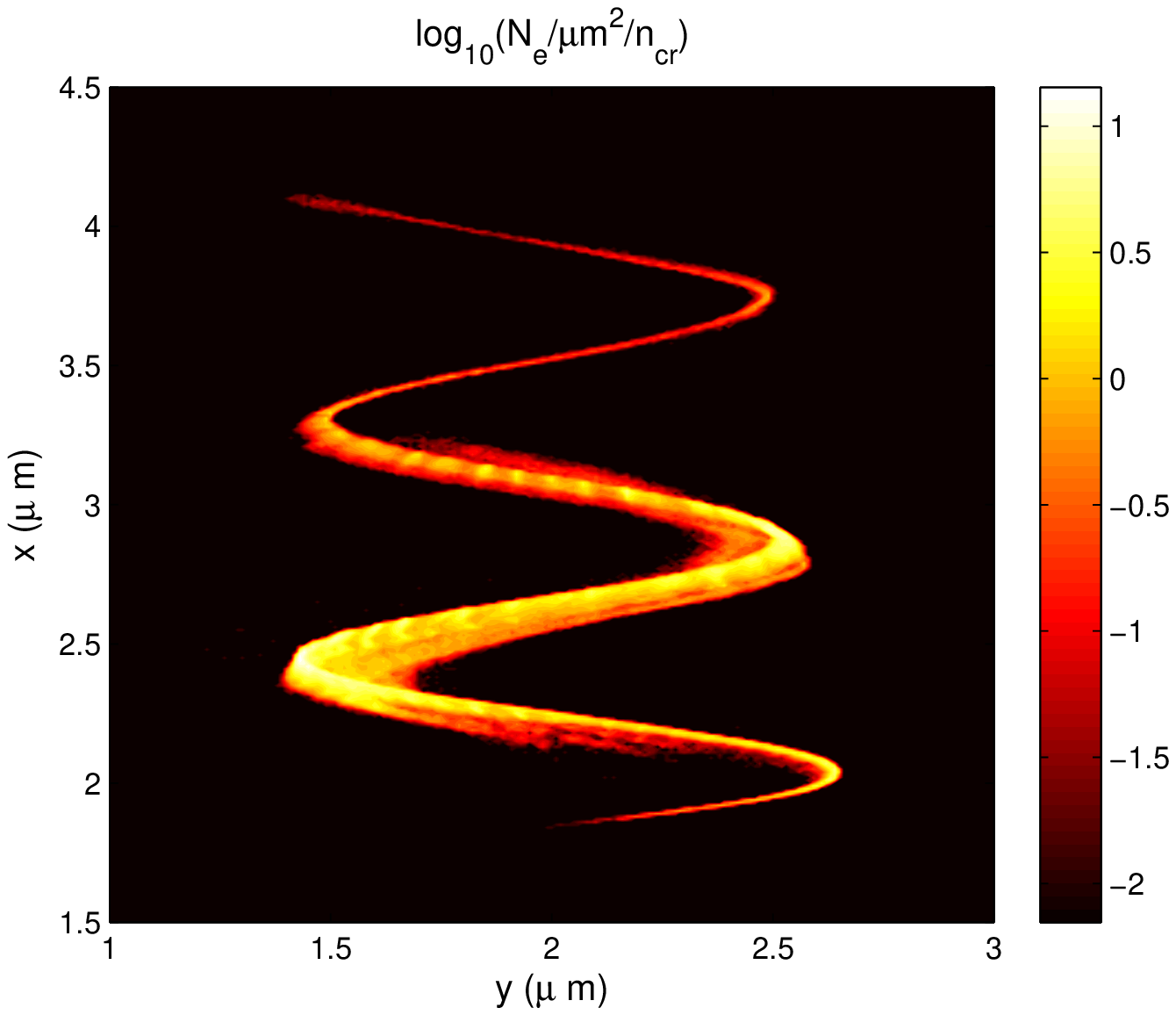}
\caption{  Electron density projected on the $xy$ plane for $\gamma_e>6$ (left) and $\gamma_e>20$ (right) in the case of cylinder target (top) and cone target (bottom) at the time instance $t=15$ fs. The dashed lines show the initial plasma surface. }
\label{eldensg6}
\end{figure}

In order to facilitate a better understanding of the electron dynamics and to ensure higher resolution, the interaction geometry is reduced to 2D and presented in the supplemental material \cite{supplem}. By analyzing the electron current and velocity distribution near the plasma surface, the source of coherent radiation could be identified in space. The 3D mechanism is the same, but  an additional angle will also define the emission angle ($\theta$) of the radiation due to the azimuthal motion of electrons. Fig.~\ref{eldensg6} shows three distinct electron populations, each traveling on different spiral path: (i) inside of the cylinder pushed forward by the laser pressure, near the inner side of the plasma wall, (ii) inside the plasma wall and (iii) outside of the plasma spreading in radial direction. The emission of the observed attopulses is related to the inner electron bunches, which have the highest density and energy (see pictures with $\gamma_e>20$). In the case of cone target the electrons form a more compact continuous spiral, which is explained by the better electron bunching observed in 2D (see supplemental material \cite{supplem}).

\begin{figure}
\includegraphics[width=40mm]{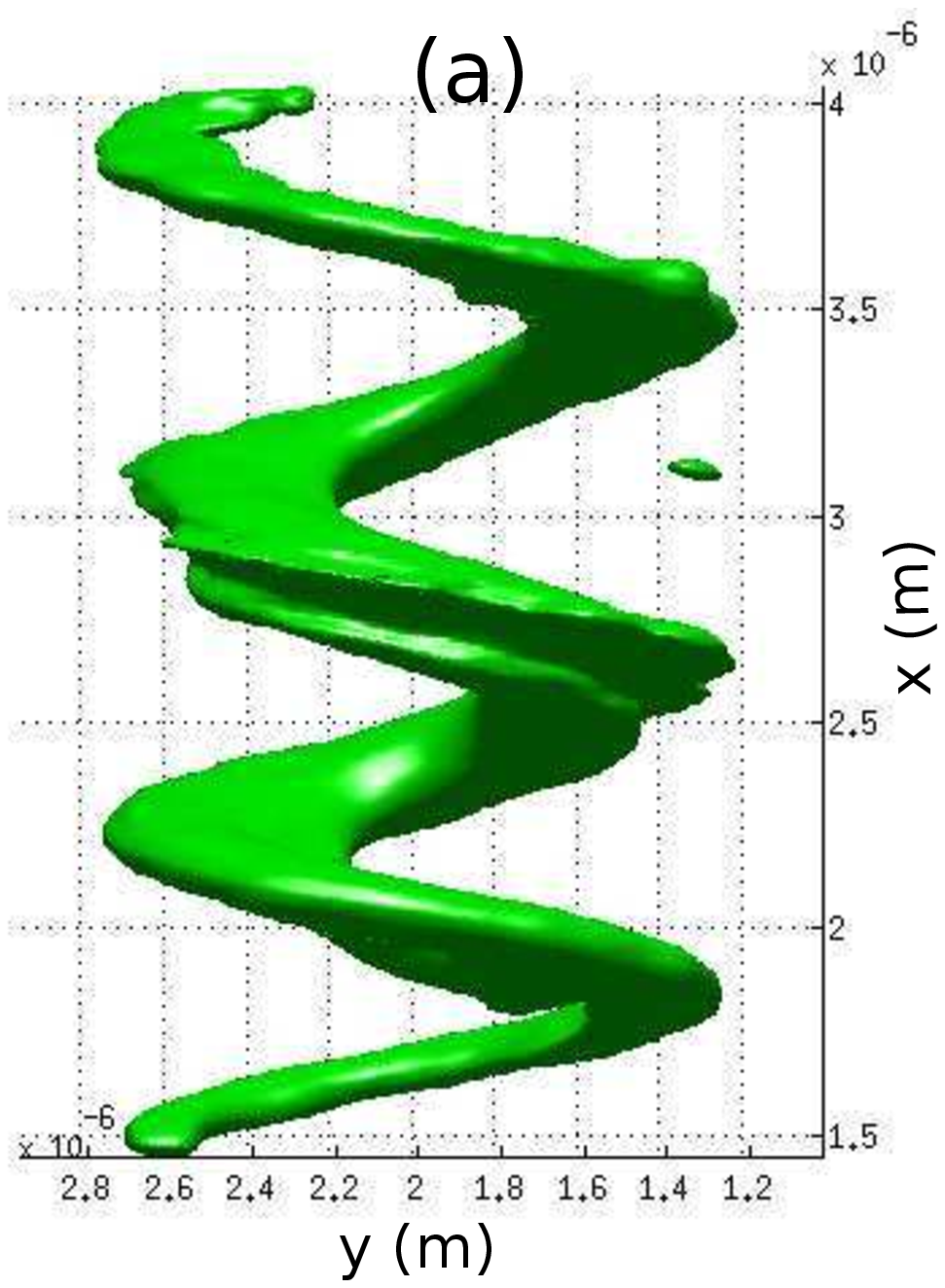}
\includegraphics[width=38mm]{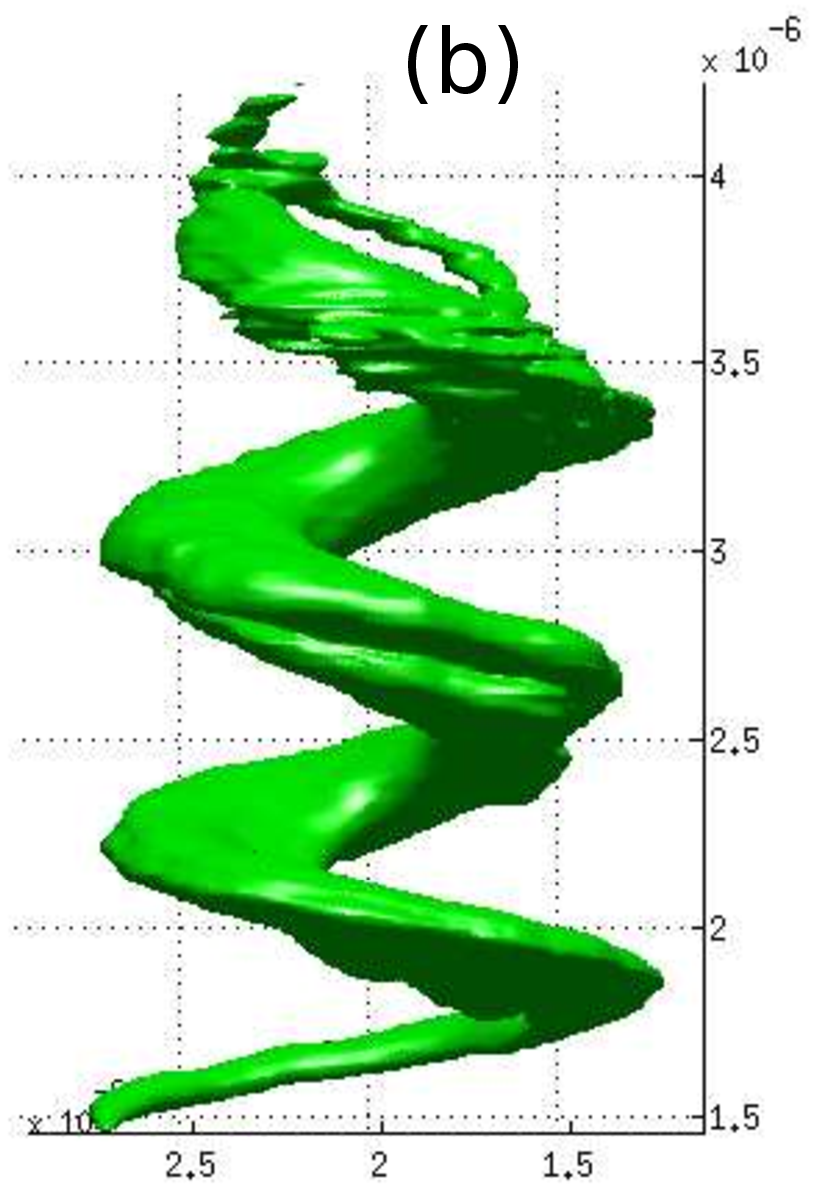}

\caption{The isosurface of electromagnetic energy density produced with cylinder (a) and cone (b) targets is shown at $t=20$ fs for the value $u=3\cdot 10^{13}$ J/m$^3$ (a) and $u=12\cdot 10^{13}$ J/m$^3$ (b). The frequencies 5 times larger than the fundamental are included. }
\label{ef_en_dens}
\end{figure}

The relativistic electrons traveling on a spiral path inside of the con-like target produce radiation according to the basic laws of synchrotron emission. Due to the relativistic longitudinal velocity the emitted field observed behind the target at a static position will be packed in a very short pulse with about a hundred of attoseconds duration. The spiral shape is preserved also in the attopulse energy distribution, which is shown in Fig. \ref{ef_en_dens}. In this figure the energy density, defined as  $u=\varepsilon_0(E_y^2+E_z^2+c^2B_y^2+c^2B_z^2)/2$, of the filtered electromagnetic field extracted from the spatial data at a given time is shown. The maximum harmonic number associated with the plasma oscillations is $\sqrt{28}\approx 5$, thus frequencies above this value are used in order to include attopulses originated from the electron bunches. For comparison, the energy density corresponding to the peak amplitude of the laser pulse is $u_L\approx 3.3\cdot 10^{15}$ J/m$^3$, which is close to the peak value measured in the focus. 

\begin{figure}[t]
\centering
\includegraphics[width=0.18\textwidth]{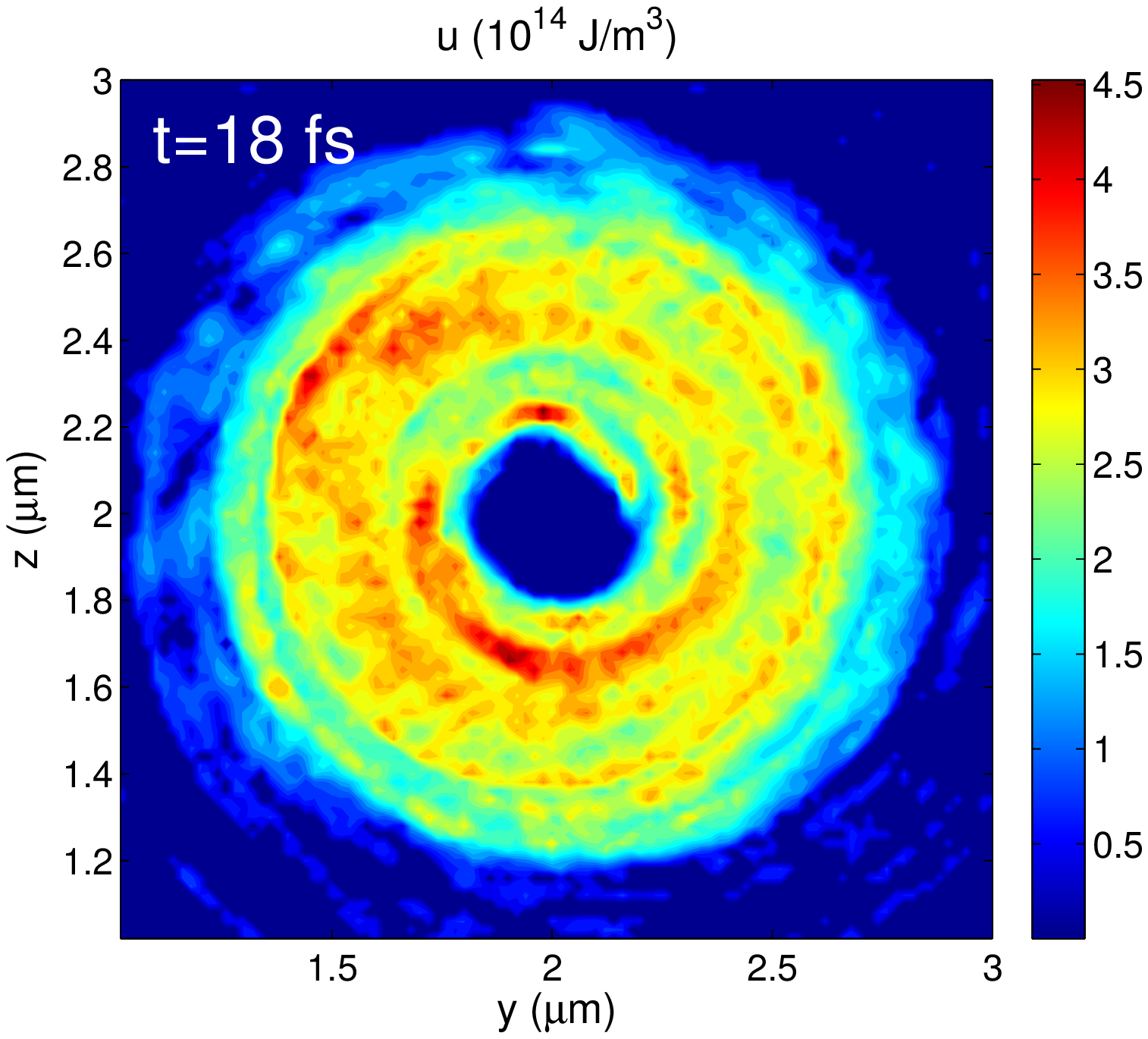} \hspace{-8mm}
\includegraphics[width=0.175\textwidth]{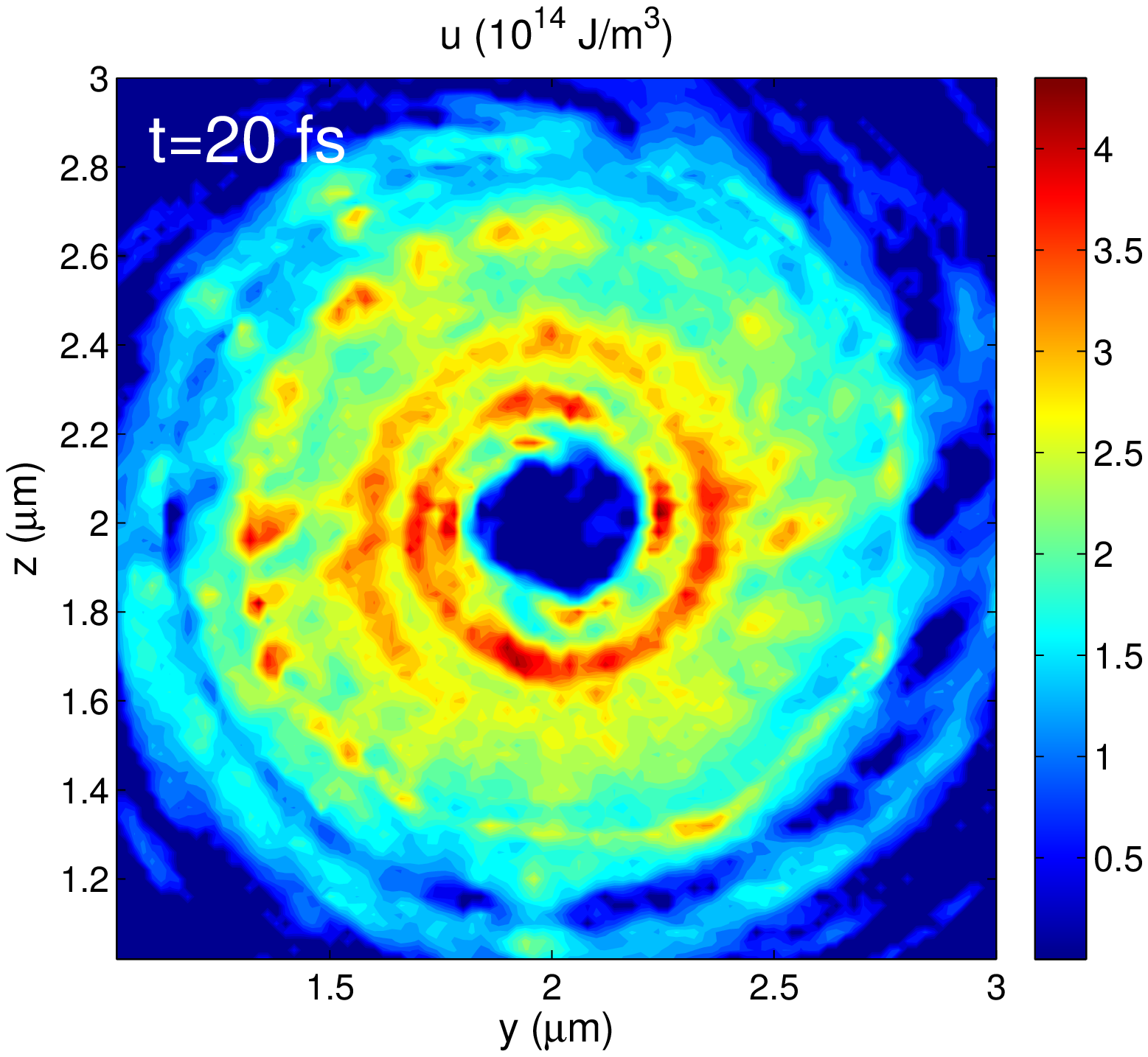} \hspace{-8mm}
\includegraphics[width=0.18\textwidth]{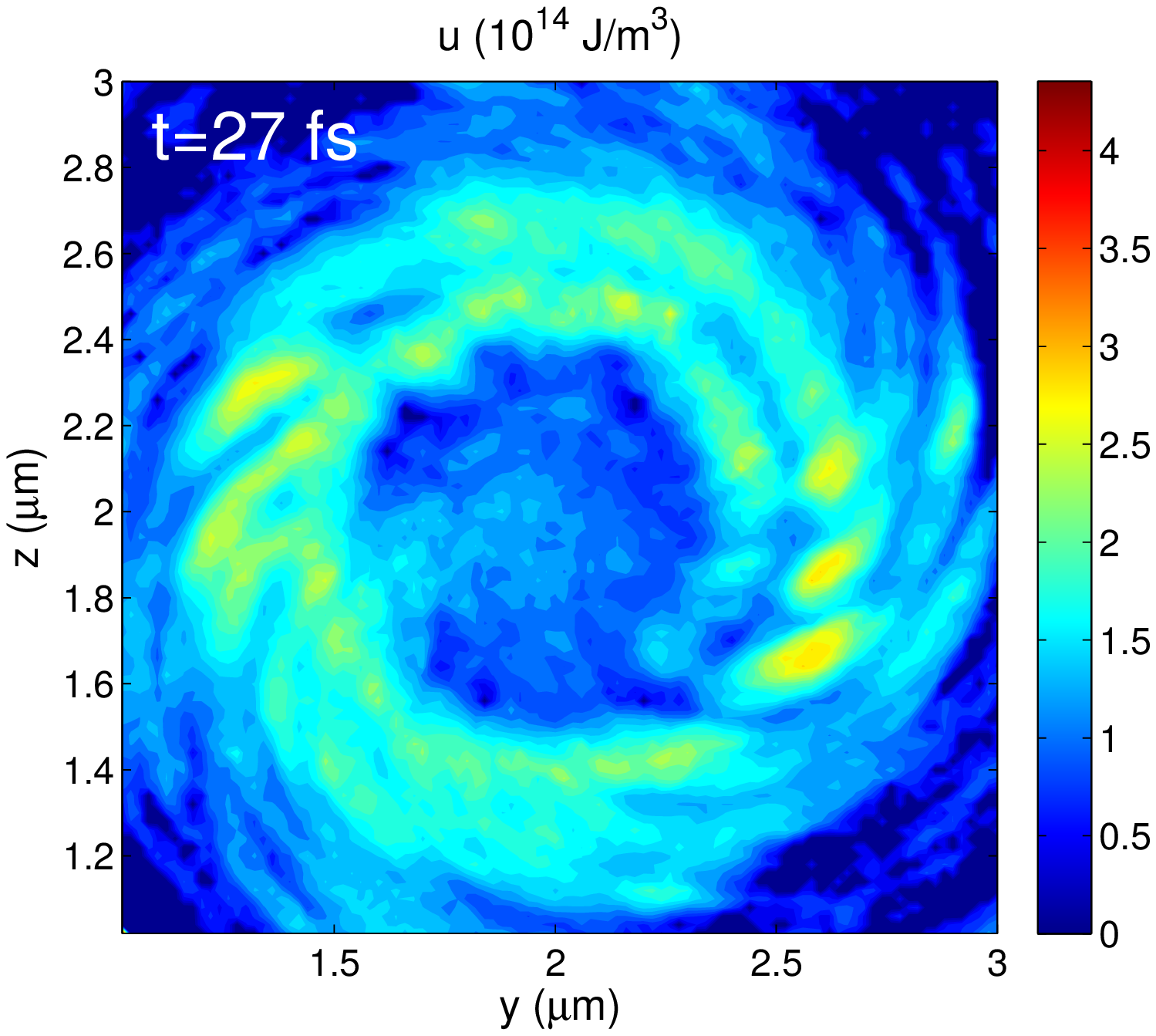}
\includegraphics[width=0.16\textwidth]{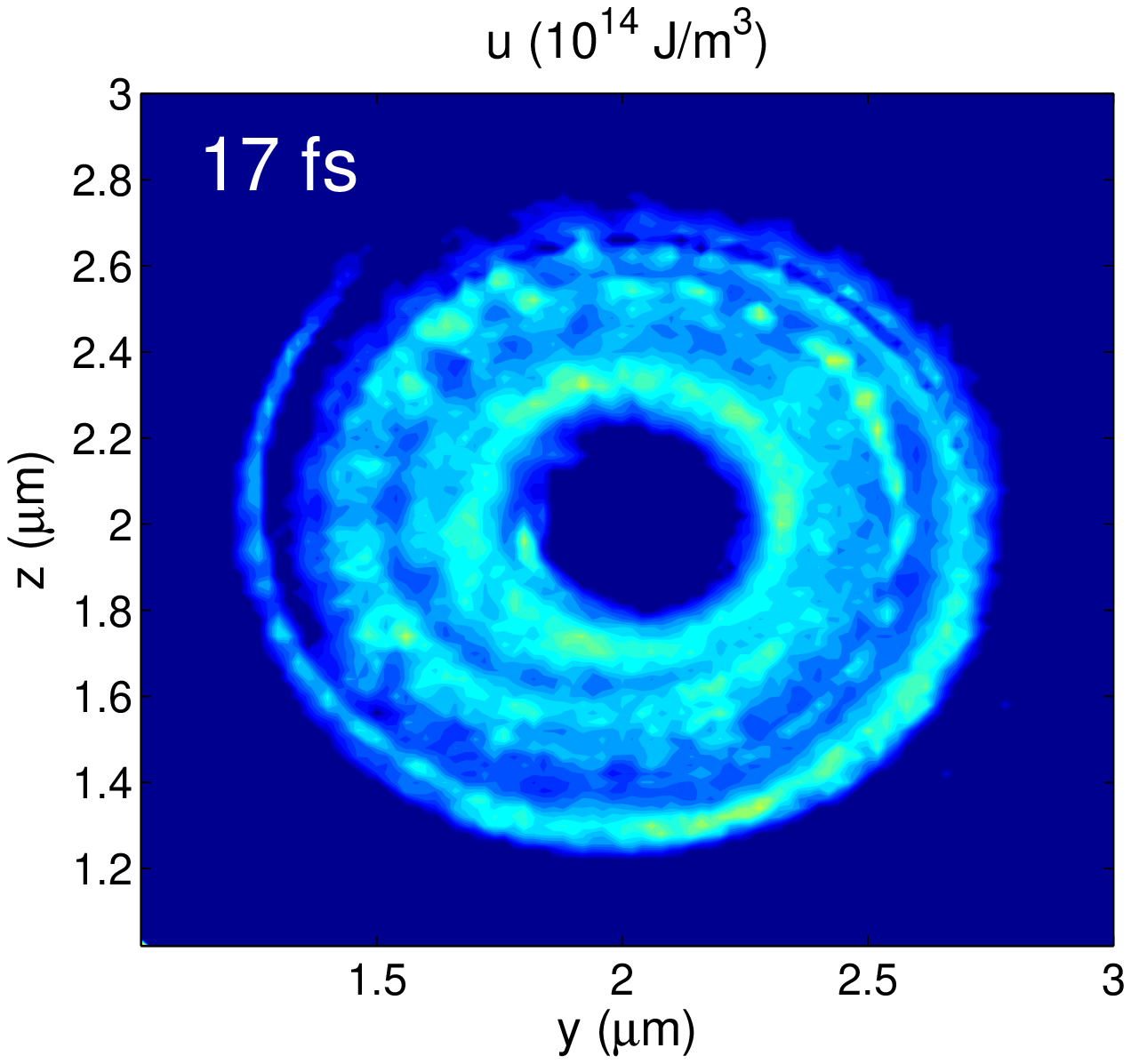} \hspace{-4mm}
\includegraphics[width=0.165\textwidth]{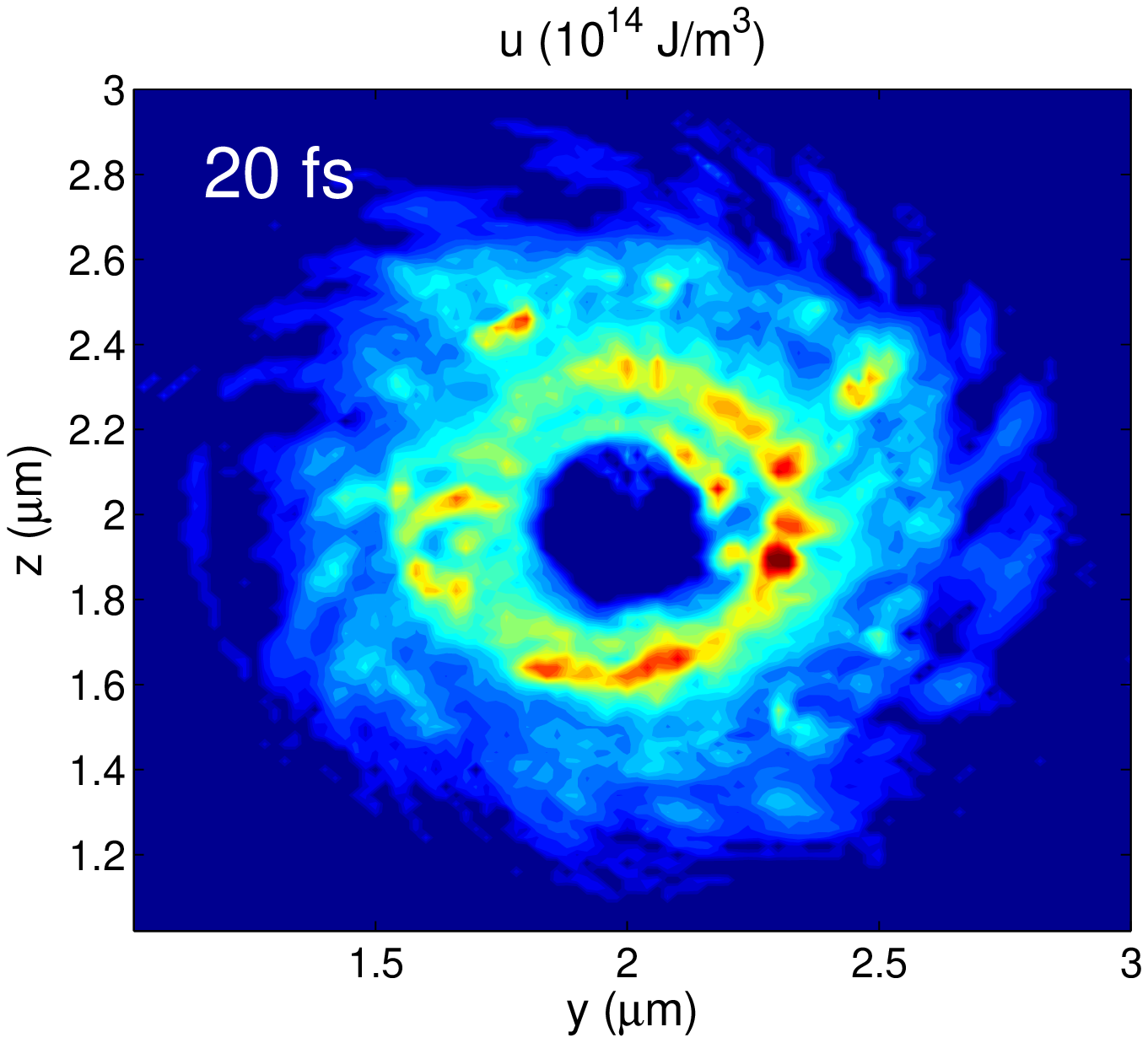} \hspace{-4mm}
\includegraphics[width=0.175\textwidth]{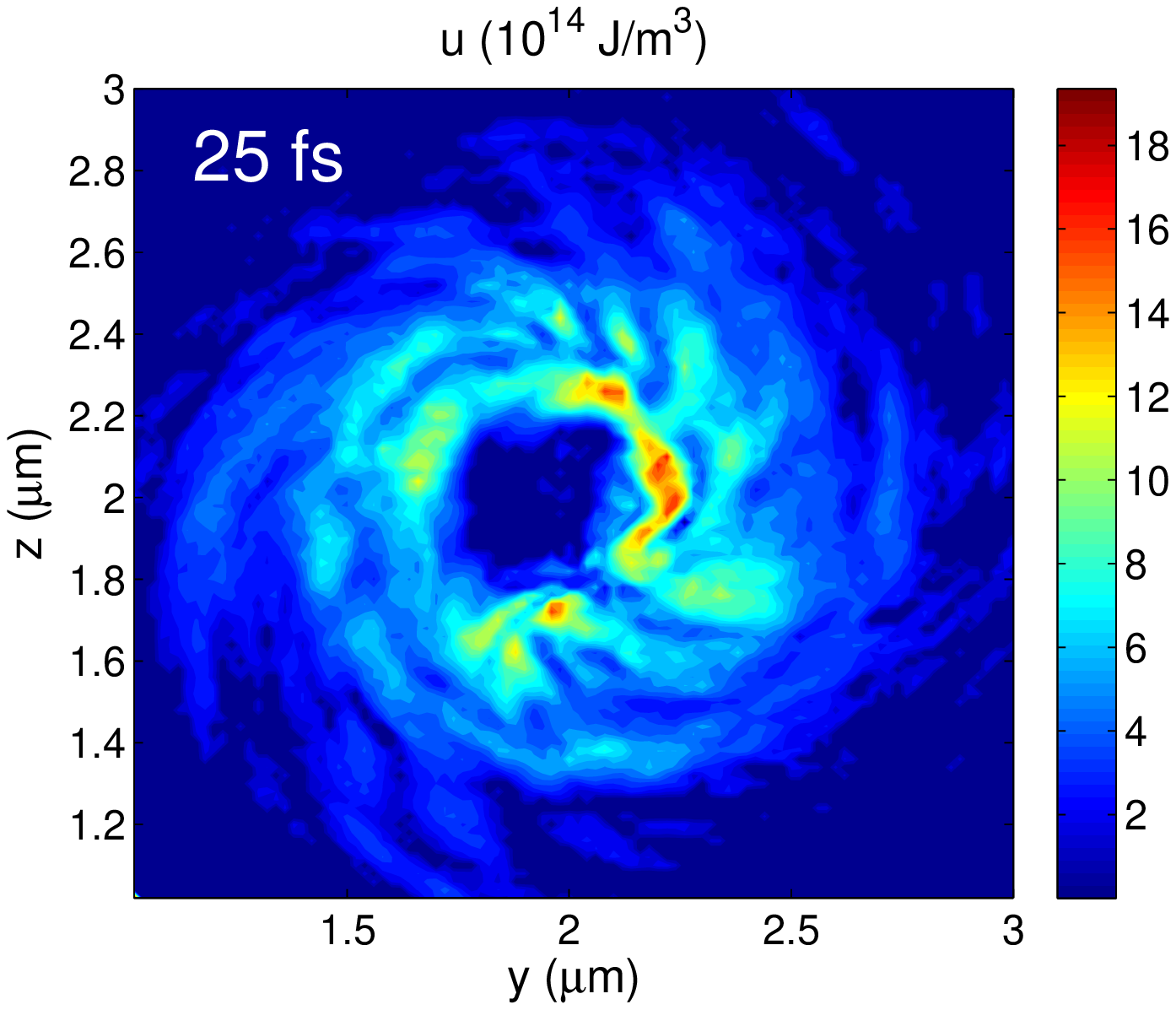}
\caption{ Averaged energy density distribution viewed from the direction of laser propagation corresponding to cylinder (top) and cone (bottom) targets. In averaging only the grid cells with $u>10^{14}$ J/m$^3$ are included. }
\label{topview}
\end{figure}

Approximately 0.4 \% and 0.9 \% of the incident pulse energy is converted into attopulses for the cylinder and cone target, respectively. This low efficiency arises because the plasma does not interact with the high intensity component of the laser pulse. If only the energy of interacting part of the Gaussian beam (between $r=0.4\,\mu$m and $0.8\,\mu$m) is considered, then the efficiency is about 3 \%. This suggests this efficiency is achievable with an annular laser beam intensity distribution, where the intensity in the center of the beam is zero (see simulation setup in \cite{supplem}). The broadening of a divergent spiral can already be seen at 25 fs in Fig.~\ref{topview}(top). However, the energy density in the focus point can reach the $u>10^{15}$ J/m$^3$ value corresponding to $I_{atto}\approx 10^{20}$ W/cm$^2$ reaching the peak intensity of the incident pulse.

The focusing effect is clearly visible in Fig.~\ref{topview}, where the transversal energy density distribution of high frequency fields is shown at different time instances. Averaging is made in the $x$ direction, which is the summation of energy density in grid cells containing $u>10^{14}$ J/m$^3$ and divided by the number of these cells. For the cone target, the energy is more concentrated in a small circle near the laser axis at $t=20$ fs which then laterally expands. This focus point is at $x \approx 3.2 \mu$m. These pictures also show that the attopulses have a ring structure with isocentric arrangement, meaning that the angular distribution of the resulting radiation is not uniform. 

The propagation direction of the attospiral defines the focusing distance, where the maximum peak intensity is expected, thus the details of the emission process has to be analyzed. For that we use the temporal evolution of the current density near the plasma surface. This method is widely used in 1D geometry in order to calculate the harmonic spectra. In order to transpose this technique into 3D, the laser-plasma interaction is assumed to be rotational symmetric, which is justified by our alignment conditions: the axis of symmetry of the target coincides with the propagation axis of the laser pulse. Thus, the current density depends on $x$ and $t$, which is equivalent with the phase of the laser pulse. The spatio-temporal distribution of the current density components is very similar to the one presented in the supplemental material \cite{supplem}.

\begin{figure}[h]
\centering
\includegraphics[width=42mm]{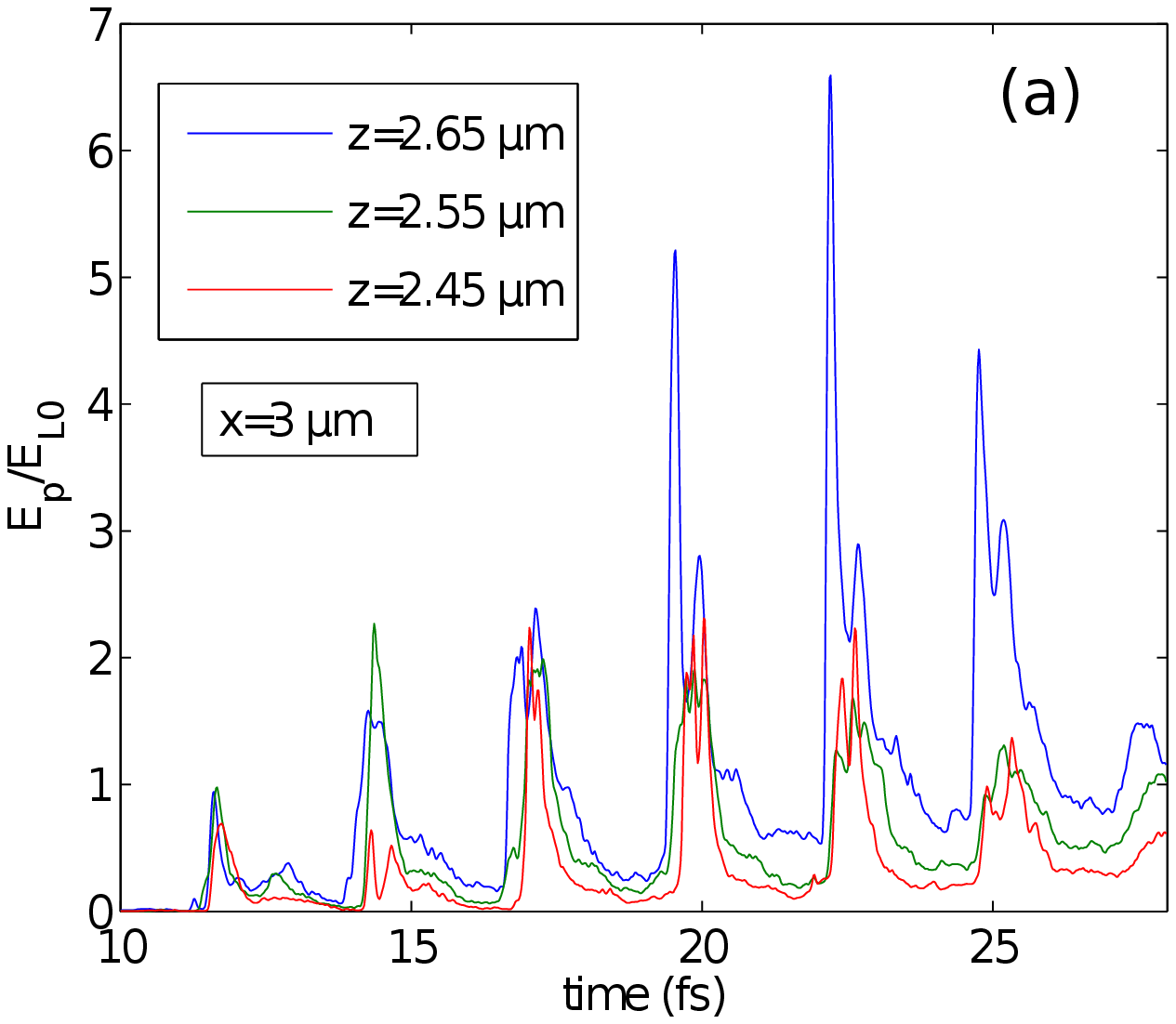}
\includegraphics[width=40mm]{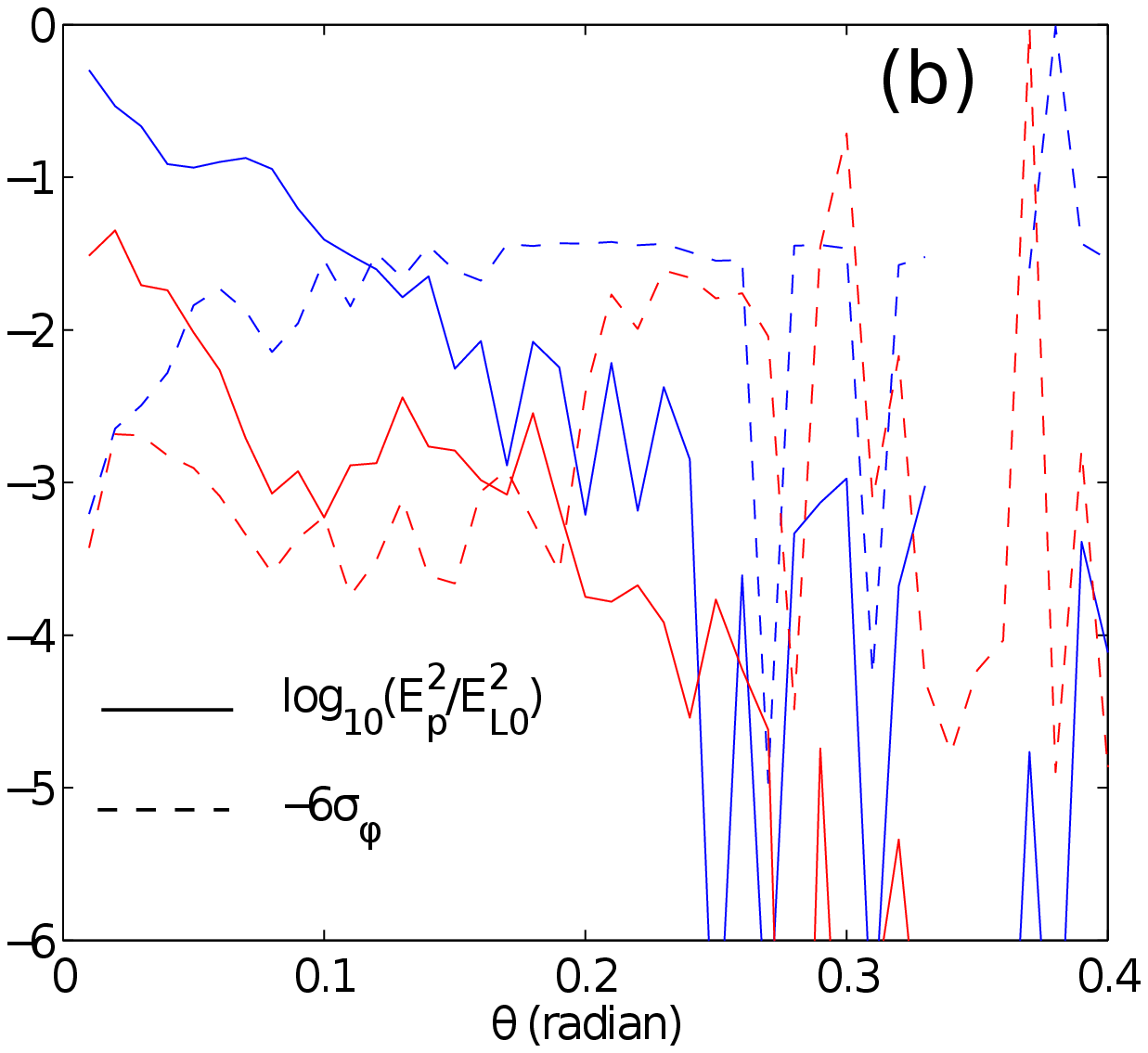}

\caption{ Electric fields measured at $x_0=3\,\mu$m using Eq. (\ref{eq:efield}) along different lines parallel with the cylinder wall (a). In all cases $y=2\,\mu$m. In (b) the angular distribution of the radiation for the strongest attopulse is shown including only frequencies higher than 5 times the laser frequency. The dashed lines show the RMS value of the frequency-dependent group delay defined by Eq. (\ref{eq:coher}). The color code is the same in both pictures.  }
\label{bunches}
\end{figure}

The transverse electric field at a given position in the laboratory frame is calculated by integrating the current density as a function of retarded time, $t'=t-(x'-x)/c$. The distance from the axis $r=\sqrt{y^2+z^2}$ is chosen arbitrarily and the mathematical form is

\begin{equation}\label{eq:efield}
E_{p}(r,x=x_0,t)= 0.5\mu_0 c^2 \int_{0}^{t} j_{p}(r,x',t') \mathbf{d}t',
\end{equation}
where $x_0$ is a chosen point where the emitted light is observed from and $j_{p}=\sqrt{j_y^2+j_z^2}$ is the transverse current density in the $yz$ plane. The measured transverse electric field from the current density is shown in Fig.~\ref{bunches}(a) at different distances from the plasma surface. The amplitude of the attopulses emitted near the plasma wall is much higher than the peak amplitude of the laser field. However, in the far-field the attopulse will have lower amplitude because of divergence and incoherence.

The $j_x$ component contains information about the emission angle of the EM wave at a given position. Ref. \cite{theory} stated that the surface normal ($e_n$) of the emitting electron layer is perpendicular to the emitted electric field, $\mathbf{e}_n\times \mathbf{E}(x,t)=0$, which, in turn, is parallel to the net current carried by the electrons, $j=\sqrt{j_x^2+j_p^2}$. From this, the emission angle can be expressed as $\theta(x,t)=\arctan(E_x/E_p)$, where the longitudinal component is calculated as $E_x=j_x \Delta t$. This component propagates in the
radial direction thus it is different in each point along a line parallel with $x$ coordinate. Knowing that the time and space data is discrete, the opening angle of the wave vector with respect to the $x$ axis can be calculated as:

\begin{equation}\label{eq:theta}
\theta(x,t)= \arctan \frac{-j_x}{\sum_{\substack{i=[1..n]}} j_p(x',t_i')},
\end{equation}
where $t_i'=i\Delta t$, $n$ is the number of time steps and $\Delta t$ (duration of one time step of simulation) is simplified.

The frequency components of the emitted radiation can be obtained by applying a Fourier transform on the calculated electric field from Eq. \ref{eq:efield}. A stable pulse structure only appears if there is a constant phase difference between consecutive frequencies. Ideally, the group delay should be constant over the whole spectrum, but it is usually not possible. Extremely short pulse can be coherently emitted from an extremely thin electron layer (nanobunch) \cite{attobunches}, which consists of relativistic electrons travelling nearly at the speed of light in the $x$ direction ($v_x\approx c, \gamma_{ex}\gg 1$). Using Eq. \ref{eq:theta}, it is possible to scan the spectrum over a wide range of emission angles.

Emission angles were calculated for every $(x,t)$ points using Eq. (\ref{eq:theta}), then the current densities in Eq. (\ref{eq:efield}) were integrated including the points where the angle is between $\theta$ and $\theta+\delta \theta$, where $\delta \theta$ can be arbitrarily small. In this way, the temporal profile of the emitted pulses is obtained for each angle and by applying the Fourier transformation, the angular dependent spectrum is obtained which is integrated from $\omega=5\omega_L$ up to a cut-off frequency (filtering) to give the attopulse energy within a given angular interval. The resultant angular distribution of the integrated power spectrum  (full lines) is shown in Fig. \ref{bunches}(b).

The Fourier transformed field data contains also the phase of the high harmonics. The degree of coherence, the standard deviation of the group delay, is used in order to characterize the emitted radiation and is defined as:

\begin{equation}\label{eq:coher}
\sigma_{\phi}(\omega)=\sqrt{< (<\tau_g(\omega)>-\tau_g(\omega))^2 >},
\end{equation}
where $\tau_g=d\phi/d\omega$. It is shown in Fig.~\ref{bunches}(b) that the majority of the coherent radiation is emitted near the plasma surface between 0.1 and 0.3 radian, where the measured energy density is relatively low. This explains the low energy conversion efficiency earlier observed and means that only a small fraction of the radiation is coherent and survives far from the surface.  Farther from the plasma surface ($z=2.45\,\mu$m), the radiation intensity is much lower and it is less coherent (red curves). The peaks in the energy distribution (Fig.~\ref{bunches}(b)) at large angles explain the ring structure observed in Fig.~\ref{topview}(top).

\begin{figure}[h]
\centering
\includegraphics[width=43mm]{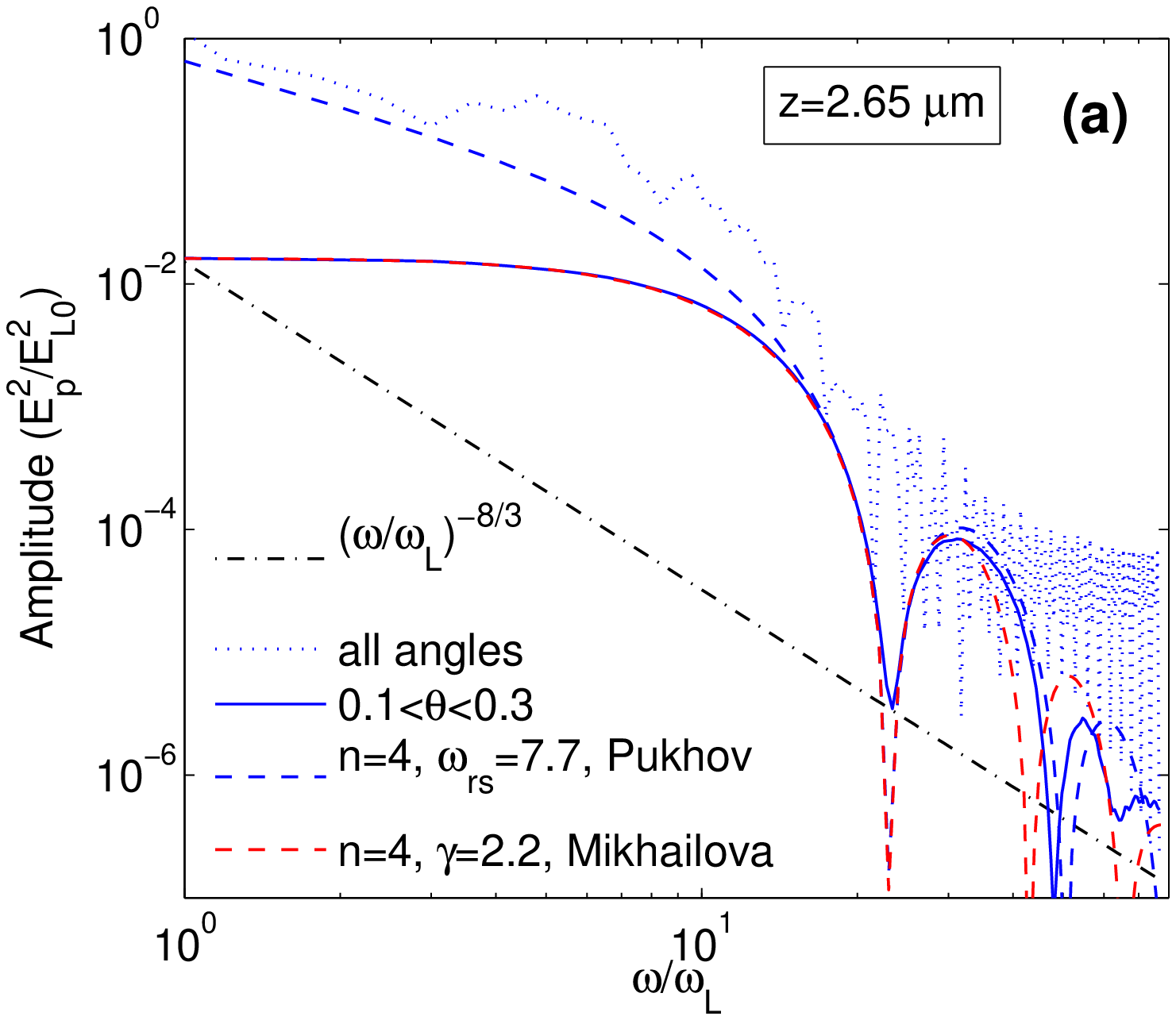} \hspace{-4mm}
\includegraphics[width=42mm]{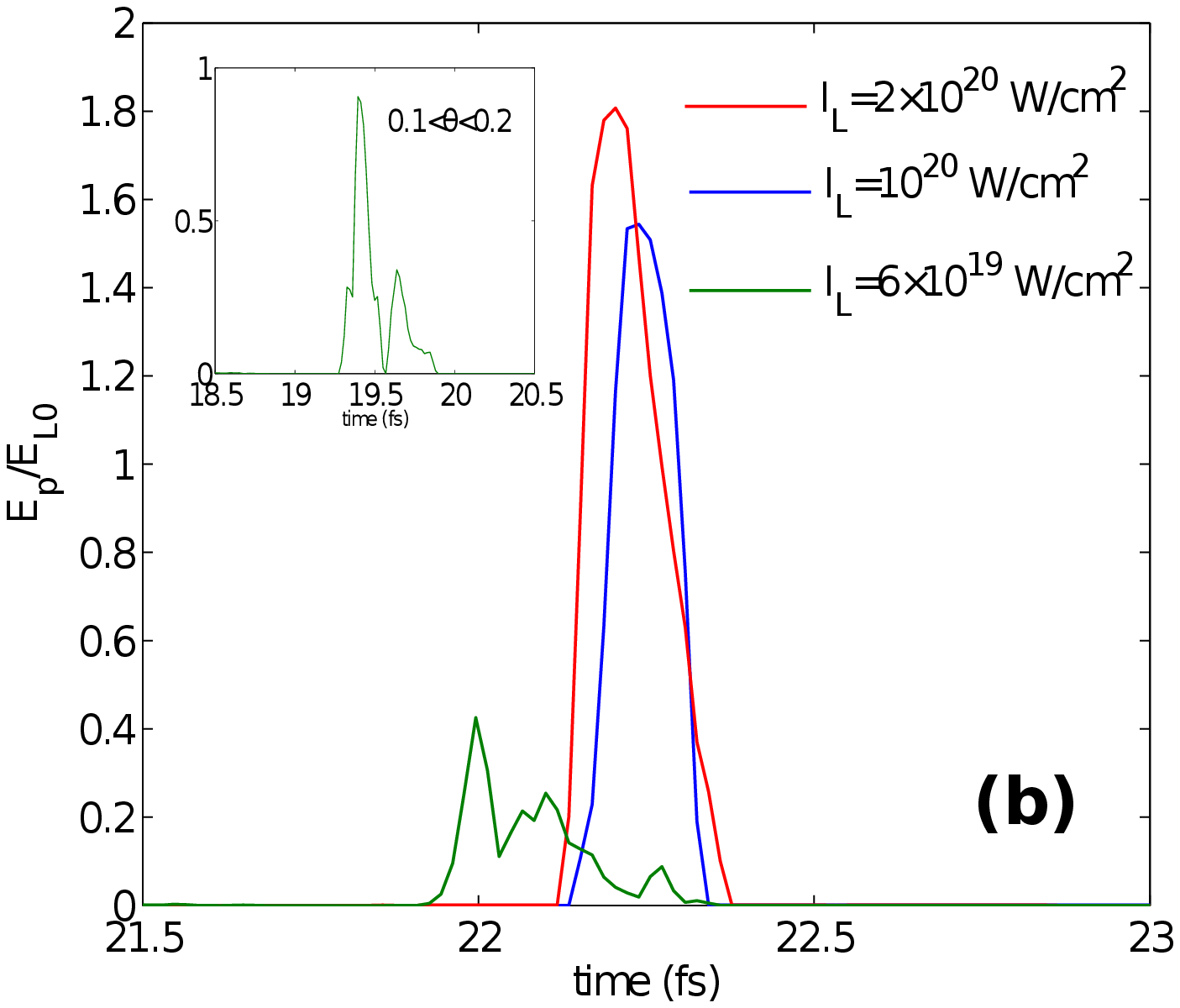}
\caption{ Frequency spectrum (a) of the strongest pulse shown in Fig.~\ref{bunches}(a) and the corresponding normalized attopulses generated within the angle of coherent radiation for different laser intensities (b). The inset in (b) shows the attopulse produced during the previous period of the laser pulse. }
\label{speccomp}
\end{figure}

The scaling of the spectral intensity with the harmonic number is very important in the generation of attopulses. Ideally, the amplitude of equally separated frequencies should be the same in order to produce high intensity and short pulses. The length of the attopulse also depends on the spectral interval, which is kept after filtering \cite{krausz2006}. 
In Fig.~\ref{speccomp}(a) the Fourier transform of the full radiation (dotted line) is compared to the spectrum emitted between 0.1 and 0.3 radians (continuous line), where the highest degree of coherence is observed. The attopulses shown in Fig.~\ref{ef_en_dens} and \ref{topview} are composed by these higher harmonics, which resemble the spectrum derived in the regime of CSE in Ref. \cite{pukhov}.

The general expression for the spectral intensity emitted by a bunch of charged particles is given by \cite{Jackson}:

\begin{equation}\label{eq:Igen}
I(\omega) \sim \left| \int \mathrm{dt} \vec{\epsilon} \times (\vec{\epsilon}\times \mathbf{J}(\mathbf{r},t)) \exp[i\omega (t-\vec{\epsilon}\cdot \mathbf{r}/c)] \right|^2,
\end{equation}
where $\vec{\epsilon}$ is the direction vector, where the radiation is observed from, and $\mathbf{r}$ is the position vector of the electron bunch. 
Since the radiation emitted in the forward direction ($x$) is of primary interest, the vector product in the above expression can be replaced by $j_p$. The argument of the exponential function can then be written as $\omega (t-x(t)/c)$. At the ultra-relativistic limit, the velocity components are defined by the direction of motion because the absolute velocity is almost constant: $v=\sqrt{v_x^2+v_{p}^2}\approx c$, where $v_x=\dot{x}(t)$ and $v_{p}$ are the longitudinal and transversal velocity, respectively. In \cite{supplem} we have obtained an expression for $v_x$, which can be approximated by a polynomial function: $v_x\approx v(1-\alpha_1 \tau^{2n})$, where $v=v_x(0)$ is the maximum velocity and $\tau$ is dimensionless. The polynomial expansion of $v_x$ shows that: $\alpha_1 \approx 6/a_0^3$ for $a_0\gg 1$. Now the transversal velocity (proportional to $j_p$) is expressed as: $v_p= \sqrt{v^2-v_x^2}\approx \sqrt{2\alpha_1}v\tau^n$. These time dependencies have been used in \cite{pop_pukhov, Pukhov2009} as well for $n=1$ ($j_p$ crosses zero) and $n=2$ ($j_p$ only touches zero), but higher values have not yet been considered.

In our case higher values of $n$ are also permitted if the electron bunch is accelerated fast enough to nearly the speed of light and propagates together with the laser pulse for a relatively long time. The final expression for the harmonic spectrum can be formulated with generalized Airy functions \cite{pukhov}, or using the Lienard-Wiechert potentials the time dependent electric field can be derived \cite{Mikhailova,supplem}, then transformed into the frequency domain to obtain the curves presented in Fig. \ref{speccomp}a. The blue dashed line is evaluated according to the numerical integration method described in \cite{pukhov}, where $\omega_{rs} \approx (\sqrt{2}\gamma)^{(2n+1)/n}(\alpha_1)^{1/2n}$ is a fitting parameter. The red dashed line is the Fourier transform of the electric field derived in \cite{supplem}, which shows a better agreement with the simulations.
 
There is a good agreement for $n=4$ and the value of $\omega_{rs}$, with $a_0\approx 4$ in the interaction region, suggests that the average $\gamma$ factor should be around 2, which is close to the average value obtained in the electron bunch. This value is confirmed by the second fitted curve (red dashed line), where $\gamma=2.2$ is used. The value of $n$ depends on laser parameters and needs to be investigated in future studies. In the supplemental material, the spectrum obtained for the cone target is presented. In that case high intensity part of the laser also interacts with the plasma providing higher $\gamma$ of electrons. The numerical fitting gives a higher values of $n$, as at higher intensities, the bunch velocity $v(1-\tau^{2n})$ stays close to $c$ for longer time.

Fig.~\ref{speccomp}(b) shows the reconstructed (inverse Fourier transform of the extracted coherent spectra) attopulses  with a duration around 100 attoseconds. Two more simulations have been performed in order to investigate the dependence of attopulse parameters on laser intensity. The plasma density is increased to 39$n_{cr}$ when higher intensity is used. With lower intensity we could find coherently emitted radiation only in the preceding laser cycle, which is shown in the inset and has a higher peak value despite of the lower intensity at that time. By looking at the spectrum in Fig. \ref{speccomp}a we can introduce an $\omega_{dr}$ harmonic number at which the first modulation starts and the intensity drops significantly. It is reasonable to consider constant spectral intensity up to $\omega_{dr}/2$, which results in attopulse duration $t_{atto}\approx (0.21/\omega_{dr})t_L$ and intensity $I_{atto}=(\omega_{dr}/2)^2I_{\omega_0}$, where $I_{\omega_0}$ is the intensity of the fundamental in the coherent radiation. We have verified numerically the dependence of $\omega_{dr}\approx 3\omega_{rs}=(3/2)a_0^2$ for $n\gg 1$, where we assumed $\gamma\approx a_0/2$. These scalings suggest that the attopulse gets shorter and more intense by increasing the laser intensity. The dependence of attopulse generation and vacuum propagation on the incidence angle is discussed in the supplemental material \cite{supplem}.

In conclusion, a new type of efficient attopulse generation and an alternative way of their focusing is presented with energy conversion efficiency as high as few percents. The peak intensity of the filtered fields in the focus can reach the value of the incident one if a cone shape target is used. Practically, a hole in a few micrometer thick target is also suitable for this process \cite{holes}. The spectral intensity measured in these simulations show a much weaker decay with the harmonic number than in the ROM model, which is very favorable for producing intense pulses even on the zeptosecond time scale \cite{zeptosecond}.

We would like to thank Ivan Konoplev and Faissal Bakkali Taheri from the John Adams Institute (Oxford) for their help and support in solving the issues regarding the parallel performance of the simulation code. The ELI-ALPS project (GOP-1.1.1.-12/B-2012-0001) is supported by the European Union and co-financed by the European Regional Development Fund.

\bibliography{Manuscript}

\end{document}